\documentclass[12pt,a4paper]{article}
\usepackage{epsfig}
\usepackage{graphicx}
\usepackage{color}
\usepackage{colortbl}
\def\pslash{\rlap{\hspace{0.02cm}/}{p}}
\def\eslash{\rlap{\hspace{0.02cm}/}{e}}
\title{The single t-quark productions via the flavor-changing  processes in the topcolor-assisted technicolor
model at the hadron colliders }
\author{Wenna Xu$^{a,b}$, Xuelei Wang$^{b,a}$\footnote{Email Address: wangxuelei@sina.com},
Zhenjun Xiao$^{a}$ \footnote{Email Address: xiaozhenjun@njnu.edu.cn} \\
\small{a: Department of Physics and Institute of Theoretical Physics, }\\
\small{ Nanjing Normal University, Nanjing, Jiangsu 210097.
P.R.China }\\
 \small {b: College of Physics and Information Engineering, }\\
\small{ Henan Normal University, Xinxiang, Henan 453007. P.R.China
}}
\begin{document}
\maketitle
\begin{abstract}
\hspace{2mm} In the framework of topcolor-assisted
technicolor(TC2) model, there exist tree-level flavor-changing
(FC) couplings which can result in the loop-level FC coupling
$tcg$. Such $tcg$ coupling can contribute significant  clues at
the forthcoming  Large Hadron Collider (LHC) experiments. In this
paper, based on the TC2 model, we study some single t-quark
production processes involving $tcg$ coupling at the Tevatron and
LHC: $pp(p\bar{p})\rightarrow t\bar{q}(q=u,d,s),tg$. We calculate
the cross sections of these processes. The results show that the
cross sections at the Tevatron are too small to observe the
signal, but at the LHC it can reach a few pb. With the high
luminosity, the LHC has considerable capability to find the single
t-quark signal produced via some FC processes involving coupling
$tcg$. On the other hand, these processes can also provide some
valuable information of the coupling $tcg$ with detailed study of
the processes and furthermore provide the reliable evidence to
test the TC2 model.
\end{abstract}

\vspace{0.5cm} \noindent {\bf PACS number(s)}: 12.60Nz, 14.80.Mz, 12.15.LK, 14.65.Ha

\newpage
\section{ Introduction}

\hspace{0.6cm}Due to the presence of the complex fundamental Higgs
scalar in the standard model(SM), the $SU(2)_{L}\bigotimes
U(1)_{Y}$ local gauge symmetry is spontaneously broken to
$U(1)_{em}$, but the presence of the Higgs scalar also leads to
the triviality and the gauge hierarchy problems. The mechanism of
electroweak symmetry breaking(EWSB) is still unclear and the
problem of large t-quark mass has not been solved yet. So many
people believe that the SM is just an effective theory, and there
might be new physics beyond the SM. A lot of works have been done
on seeking new physics, and many new physics models have been put
forward.

On the experimental aspect, the search for the Higgs boson
continues at the upgraded $p\bar{p}$ collider Tevatron which is
now engaged in RUN II, and followed in the near future by the $pp$
collider LHC with the c.m. energy of 14 TeV. On the other hand,
since the Higgs boson predicted in the SM has not been found and
the EWSB mechanism is unclear, the other primary goal of such
hadron colliders is to probe the hint of the new physics beyond
the SM. The Tevatron II has the significant potential to discover
a light Higgs boson with mass up to 130 GeV\cite{Tevatron} in the
SM or the minimal supersymmetric standard model (MSSM). However,
it will have very little capability to determine the underlying
model that governs the EWSB if a Higgs boson or Higgs-like
candidate is observed. The LHC will have a considerable capability
to discover and measure almost all the quantum properties of a SM
Higgs boson with any mass or several of the MSSM Higgs bosons over
the entire MSSM parameter space\cite{LHC}. Furthermore, the LHC
also has the capability to detect the signal of the new heavy
Higgs-like particles in the new physics models related to the
EWSB, such as the top-pions in the TC2 model. Therefore, the LHC
will open a unique window to test the SM or the new physics
models, furthermore, explore the EWSB.

To completely avoid the problems of triviality and hierarchy
arising from the elementary Higgs field in the SM, people have
proposed various kinds of dynamical EWSB models. The
technicolor(TC) model is introduced by Weinberg and
Susskind\cite{TC} and such model offers a new insight into
possible mechanisms of the EWSB. As late as 1990s, people arrived
at a viable model in which topcolor is introduced and t-quark
acquires a dynamical mass through the topcolor interaction. Such
model is called topcolor-assisted technicolor(TC2)
model\cite{TC2}.

The TC2 model offers a new possible EWSB mechanism and
successfully solves the problems of triviality and gauge
hierarchy. It is consistent with the experimental limits and
predicts the rich phenomenologies that may be accessible to the
colliders. In this model, the EWSB is mainly driven by the
technicolor interaction. The extended technicolor(ETC) gives
contributions to the masses of ordinary quarks and leptons, and it
also gives a very small portion of t-quark mass $\varepsilon
m_{t}$($0.03\leq \varepsilon\leq 0.1$)\cite{TC2}. The topcolor
interaction gives the main part of t-quark mass
$(1-\varepsilon)m_{t}$, and makes a small contribution to the
EWSB. The TC2 model predicts some new particles at several hundred
GeV scale: three top-pions( $\Pi_{t}^{\pm}$, $\Pi_{t}^{0}$) and a
top-higgs boson($h_t$). These particles are regarded as the
typical feature of the TC2 model. Another feature of the TC2 model
is that the topcolor interaction is non-universal, i.e., topcolor
treats the third family differently from the first and second
families. So, the GIM symmetry is violated which results in the
tree-level FC couplings when one writes the interactions in the
quark mass eigen-basis\cite{He}. Such FC feature also exists in
other new physics models(such as the MSSM or the two-Higgs-doublet
model (2HDM)).

The SM does not contain the tree-level flavor-changing neutral
currents(FCNC), though it can occur at higher order through
radiative corrections. Because of the loop suppression, these SM
FCNC effects are hardly to be observed, and so large FCNC provides
a unique window into the new physics. On the other hand, since the
discovery of t-quark at the Tevatron \cite{t-discover}, there have
been considerable interests in exploring the properties of
t-quark. Its unusually large mass close to the EWSB scale also
makes it a good candidate for probing the new physics. So, people
pay much attention to the processes involving t-quark and the FC
couplings, and a lot of studies about the t-quark rare decay
\cite{TCV-2HDM,TCV-MSSM,TCV-IN,TCV-TC} and  single t-quark
production processes involving the FC couplings
\cite{He,TC2-TX,TC2-TC,MSSM-TX,IN-TX,IN-TC} have been done in
recent years.

In this paper, we systematically study some single t-quark
production processes involving one loop-level FC coupling $tcg$,
i.e., $c\bar{q}\rightarrow t\bar{q}(q=u,d,s), cg\rightarrow tg$.
Some important FC single t-quark production processes, such as
$gg(q\bar{q})\rightarrow t\bar{c}$ process, have been studied in
the references \cite{TC2-TC} in the framework of the TC2 model,
the results show that the significant contributions of the FC
couplings in the TC2 model make this single t-quark production
process detectable, which opens an ideal window to test the TC2
model. Our study in this paper can provide some useful
compensatory information about FC couplings in the TC2 model and
give the valuable theoretical suggestion to probe the TC2 model at
hadron collider experiments.

The rest parts of this paper are organized as follows. In section
two and section three, we present our detailed calculation and
numerical results of the cross sections for the $t\bar{q},tg$
productions, respectively. The discussions and conclusions are
given in the final section.

\section{The $t\bar{q}$ productions}

\hspace{0.6cm}As it is known, the topcolor interaction is
non-universal which can result in the significant tree-level FC
couplings of the top-pions and top-higgs to the quarks when one
writes the interaction in the quark mass eigen-basis. The
couplings of the top-pions and top-higgs to the quarks can be
written as  \cite{He}
 \begin{eqnarray}
 \cal L&=&\frac{m_{t}}{\upsilon_{w}}
 \tan\beta
 [iK_{UR}^{tt}K_{UL}^{tt*}\overline{t_{L}}t_{R}\Pi_{t}^{0}
 +\sqrt{2}K_{UR}^{tt*}K_{DL}^{bb}\overline{t_{R}}b_{L}\Pi_{t}^{+}  \nonumber\\
 &&\hspace*{1.5cm}+iK_{UR}^{tc}K_{UL}^{tt*}\overline{t_{L}}c_{R}\Pi_{t}^{0}
 +\sqrt{2}K_{UR}^{tc*}K_{DL}^{bb}\overline{c_{R}}b_{L}\Pi_{t}^{+}   \nonumber\\
 &&\hspace*{1.5cm}+i\frac{m_{b}^{*}}{m_{t}}\overline{b_{L}}b_{R}\Pi_{t}^{0}
+K_{UR}^{tt}K_{UL}^{tt*}\overline{t_L}t_Rh_{t}^{0}+K_{UR}^{tc}K_{UL}^{tt*}\overline{t_{L}}c_{R}h_{t}^{0}
 +h.c.].
 \end{eqnarray}
Where $\tan\beta=\sqrt{(\upsilon_{w}/\upsilon_{t})^2-1}$,
\hspace{0.5cm}$\upsilon_{t}\approx 60-100$ GeV is the top-pion
decay constant, $\upsilon_{w}=246$ GeV is the EWSB scale,
$K^{ij}_{U,D}$ are the matrix elements of the unitary matrix
$K_{U,D}$, from which the Cabibbo-Kobayashi-Maskawa (CKM) matrix
can be derived as $V=K^{-1}_{UL}K_{DL}$. Their values can be
written as
 \begin{eqnarray*}
  \hspace*{1cm}K^{tt}_{UL}=K^{bb}_{DL}\approx 1,
  \hspace{1.5cm}K^{tt}_{UR}=1-\varepsilon,
  \hspace*{1.5cm}K^{tc}_{UR}=\sqrt{2\varepsilon-\varepsilon^{2}}.
  \end{eqnarray*}
The mass $m_{b}^{*}$ is a part of b-quark mass which is induced by the
instanton, and it can be estimated as \cite{TC2}
\begin{eqnarray*}
  \hspace*{1cm}m_{b}^{*}=\frac{3\kappa m_{t}}{8\pi^{2}}\sim
  6.6\kappa GeV,
  \end{eqnarray*}
where we generally expect $\kappa\sim1$ to $10^{-1}$ as in QCD.

\begin{figure}[thb]
\begin{center}
\vspace{1cm}
\includegraphics [scale=0.7] {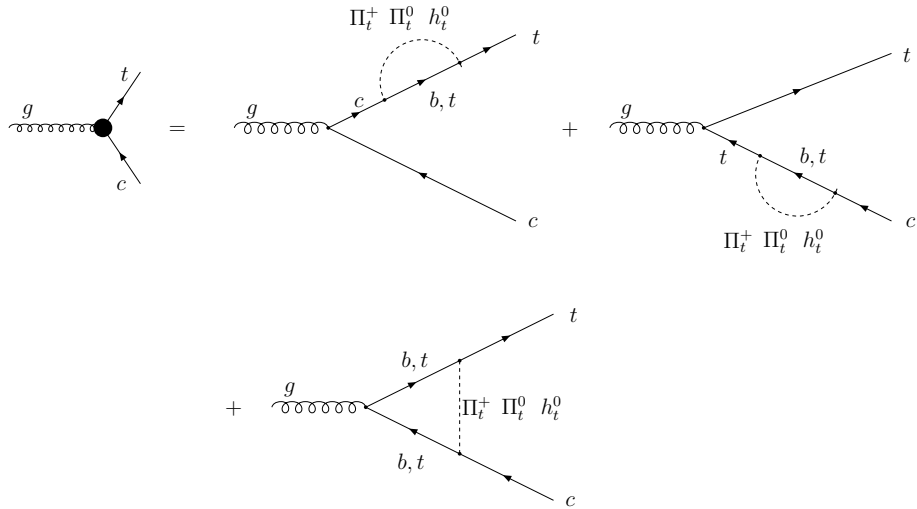}
\caption{The Feynman diagrams of the one-loop FC coupling $tcg$.}
\label{fig:fig1}
\end{center}
\end{figure}

\begin{figure}[thb]
\begin{center}
\includegraphics [scale=0.7] {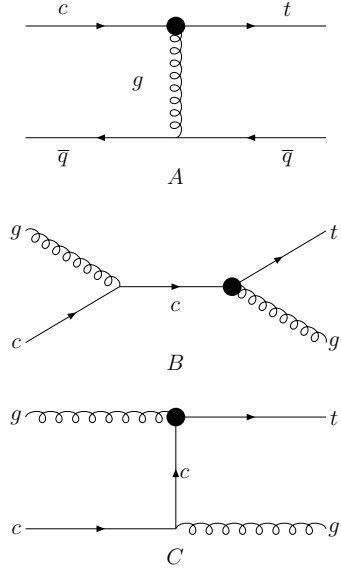}
\caption{ The Feynman diagrams of the single t-quark production
processes involving the $tcg$ coupling.} \label{fig:fig2}
\end{center}
\end{figure}

As we can see above, in the TC2 model, there exist FC couplings
$\Pi_{t}^{0}t\bar{c}$, $\Pi_t^+c\bar{b}$ and $h^0_t t\bar{c}$ at
tree-level. These tree level FC couplings can result in the
loop-level FC coupling $tcg$ as shown in Fig.1. Such FC coupling
$tcg$ in the TC2 model would make the significant contribution to
some processes. For example, it can enhance the SM branching
ratios of the rare t-quark decay $t\rightarrow cg$ by as much as 7
orders of magnitude \cite{TCV-TC}. Such FC coupling can also
significantly enhance the cross section of $t\bar{c}$ production
at the hadron colliders to make the $t\bar{c}$ signal observable
\cite{TC2-TC}. With this FC coupling, single t-quark associated
with a light quark q(q=u,d,s) can also be produced at the hadron
colliders, i.e., $c\bar{q}\rightarrow t\bar{q}$, which will
complement the information of $tcg$ coupling. These processes can
be realized via the t-channel $c-\bar{q}$ collision and the
relative Feynman diagrams are given in Fig.2(A). By calculating
the loop-level $tcg$ coupling, we can easily write the production
amplitudes of the processes. Each figure in Fig.1 actually
contains the ultraviolet divergence. Because there is no
corresponding tree-level $tcg$ coupling to absorb these
divergences, the divergences just cancel each other and the total
result is finite as it should be. The $tcg$ coupling can be
expressed in terms of two-point and three-point standard
functions. The production amplitudes of $c\bar{q}\rightarrow
t\bar{q}$ can be written as
\begin{eqnarray}
M_{A}&=&\frac{im_{t}^{2}tan^{2}\beta}{16\pi^{2}\upsilon_{w}^{2}}g_{s}^{2}(1-\varepsilon)
\sqrt{2\varepsilon-\varepsilon^{2}}(T_{ij}^{a}T_{kl}^{a})G(p_{3}-p_{1},0)\nonumber\\
&&\cdot \left\{ \left \{ 2B_{1}(-p_{3},m_{b},M_{\Pi_{t}})+B_{1}(-p_{3},m_{t},M_{\Pi_{t}})+B_{1}(-p_{3},m_{t},M_{h_{t}})
\right.\right. \nonumber\\
&&\left. \left. + B_{0}(-p_{3},m_{t},M_{\Pi_{t}})-B_{0}(-p_{3},m_{t},M_{h_{t}})
-B_{0}(-p_{1},m_{t},M_{\Pi_{t}})+B_{0}(-p_{1},m_{t},M_{h_{t}})\right. \right. \nonumber\\
&& \left. \left.  +m_{t}^{2}\left [ 2(C_{21}+C_{11})+C'_{21}+C'_{11}+C^*_{21}+C^*_{11}-2C^*_{0}\right ]\right.\right.
\nonumber\\
&& \left. \left. -2m_{b}^{2}C_{0}+4C_{24}+2C'_{24}+2C_{24}^{*} \right \}
\cdot \overline{u}_{t}(p_{3})\gamma^{\mu}Ru_{c}(p_{1})\overline{v}_{q}(p_{2})\gamma_{\mu}v_{q}(p_{4})\right.\nonumber\\
&&\left. -2m_{t}\left [2(C_{21}+C_{11})+C'_{21}+2C'_{11}+C'_{0}+C^*_{21}+C^*_{0}\right ]
\cdot \overline{u}_{t}(p_{3})Ru_{c}(p_{1})\overline{v}_{q}(p_{2})\pslash_{3}v_{q}(p_{4})\right.\nonumber\\
&&\left. +\left [2(C_{23}+C_{12})+C'_{23}+C'_{12}+C^*_{23}+C^*_{12}\right ]
\cdot \overline{u}_{t}(p_{3})\pslash_{1}\gamma^{\mu}\pslash_{3}Ru_{c}(p_{1})\overline{v}_{q}(p_{2})\gamma_{\mu}v_{q}(p_{4})
\right.\nonumber\\
&&\left. +2m_{t}(C'_{12}-C^*_{12})\overline{u}_{t}(p_{3})Ru_{c}(p_{1})\overline{v}_{q}(p_{2})\pslash_{1}v_{q}(p_{4})
\right \},
\end{eqnarray}
where $G(p,m)=1/(p^{2}-m^{2})$ is the propagator of the particle,
$R=(1+\gamma_5)/2$. In production amplitude $M_A$, the three-point
standard functions are defined as
$$C_{ij}=C_{ij}(-p_{3},p_{1},m_{b},M_{\Pi_{t}},m_{b}),$$
$$C'_{ij}=C_{ij}(-p_{3},p_{1},m_{t},M_{\Pi_{t}},m_{t}),$$
$$C^*_{ij}=C_{ij}(-p_{3},p_{1},m_{t},M_{h_{t}},m_{t}).$$
Here, we have ignored the mass difference between the charged top-pion and
the neutral top-pion.

Using above amplitudes, it is straightforward to calculate the cross sections
$\hat{\sigma}^{t\bar{q}}(\hat{s})$ for the subprocess
$c\bar{q}\rightarrow t\bar{q}$. The hadronic cross sections
$\sigma^{t\bar{q}}(s)$ can be obtained by folding
$\hat{\sigma}^{t\bar{q}}(\hat{s})$ with the parton distribution
functions. In the calculation of the cross sections, instead of
calculating the square of the production amplitudes analytically,
we calculate the amplitudes numerically by using the method of
Ref~.\cite{HZ}. This greatly simplifies our calculations.

\begin{figure}[thb]
\begin{center}
\includegraphics[scale=0.8] {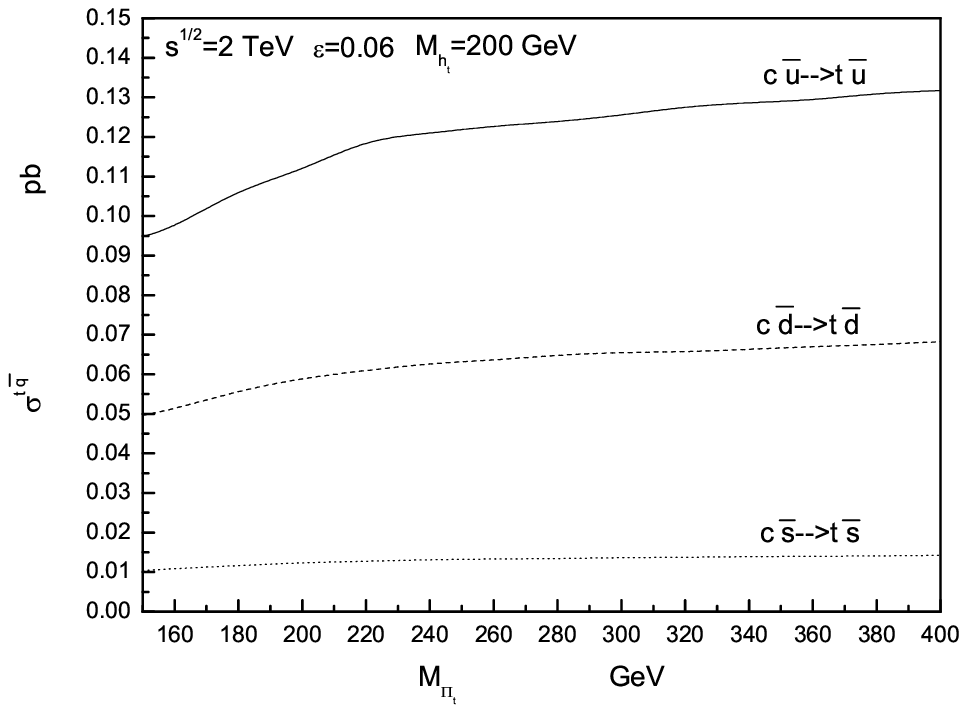}
\includegraphics[scale=0.8] {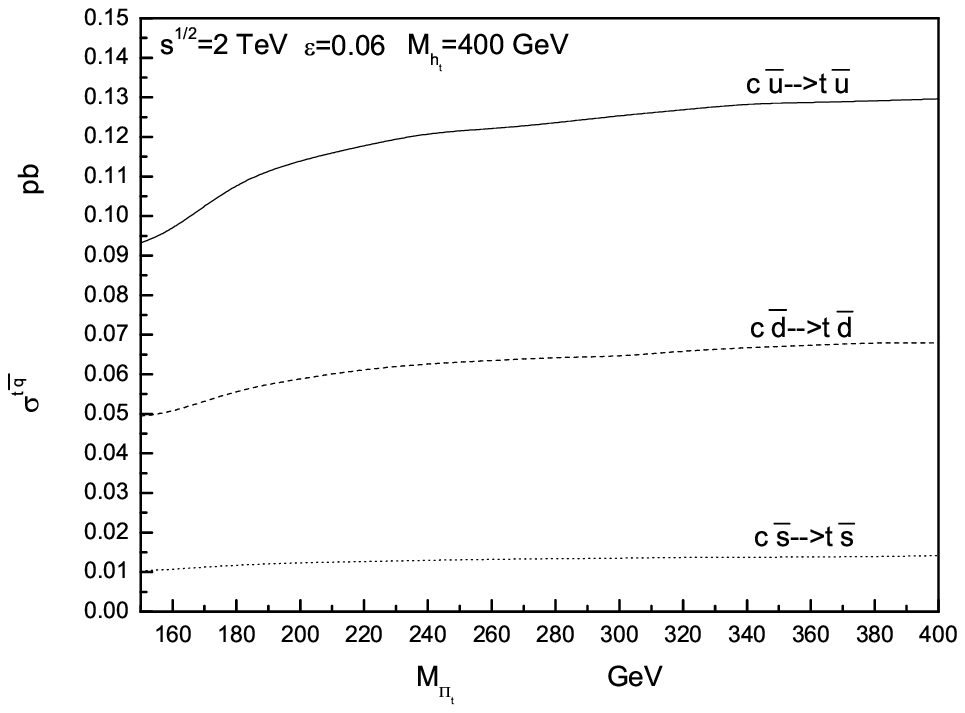}
\caption{The hadronic cross sections of $t\bar{q}$ productions as
a function of $M_{\Pi_t}$ at the Tevatron, with $M_{h_t}=200, 400$
GeV, respectively.} \label{fig:fig3}
\end{center}
\end{figure}

\begin{figure}[thb]
\begin{center}
\includegraphics [scale=0.8] {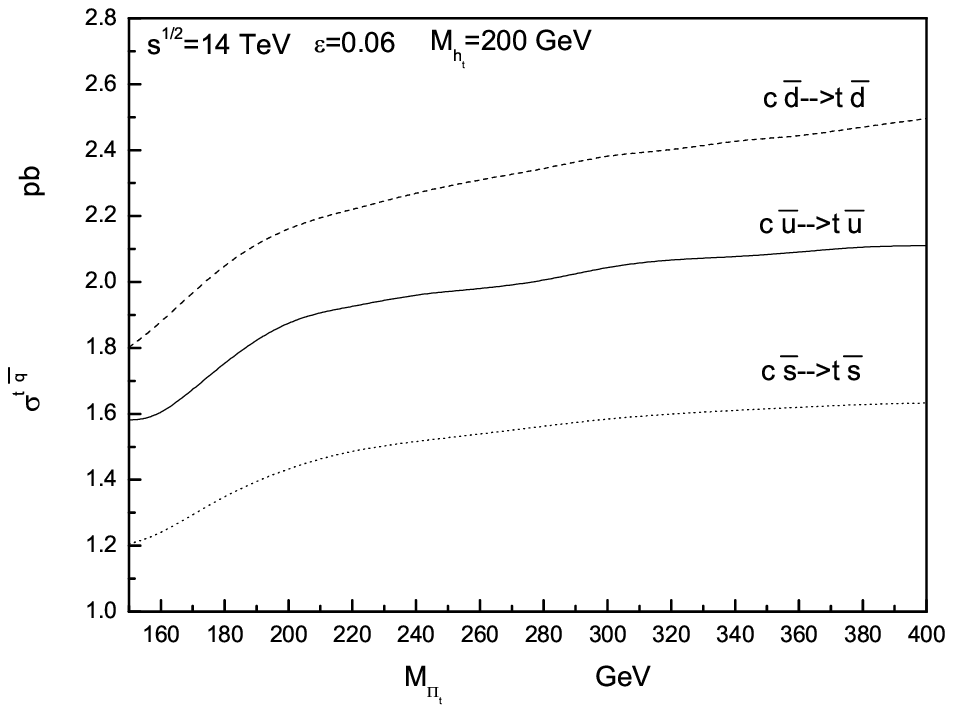}
\includegraphics [scale=0.8] {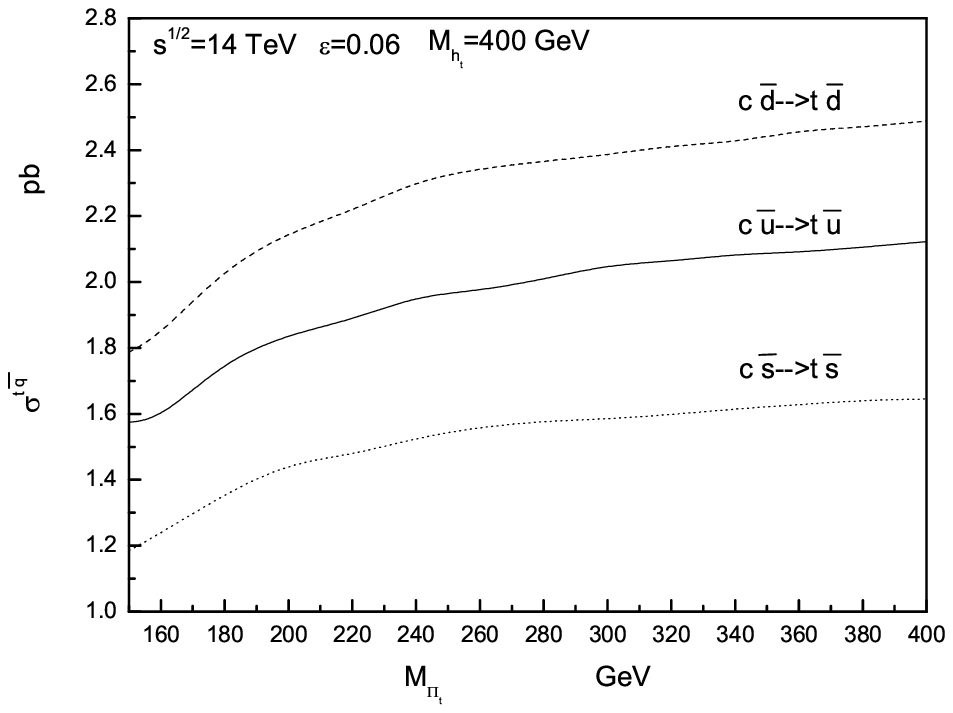}
\caption{The hadronic cross sections of the $t\bar{q}$ productions
as a function of $M_{\Pi_t}$ at the LHC, with $M_{h_t}=200, 400$
GeV, respectively.} \label{fig:fig4}
\end{center}
\end{figure}

To obtain the numerical results of the cross sections, we take
$m_{t}=174$ GeV, $m_{b}=4.7$ GeV, $m_{u}=2$ MeV, $m_{d}=5$ MeV,
$m_{s}=95$ MeV\cite{data}, $\upsilon_{w}=246$ GeV,
$\upsilon_{t}=60$ GeV. The strong coupling constant,
$g_{s}=2\sqrt{\pi\alpha_{s}}$, can be obtained from the one-loop
evolution formula at the energy of the Tevatron and the LHC,
respectively. There are three free parameters involved in the
production amplitudes: the top-pion mass $M_{\Pi_t}$(We have
ignored the mass difference between the neutral top-pion and
charged top-pion), top-higgs mass $M_{h_{t}}$ and parameter
$\varepsilon$. In order to see the influence of these parameters
on the cross sections, we take $M_{\Pi_t}$ to vary in the range
$150 GeV\leq M_{\Pi_t}\leq400$ GeV, $\varepsilon=0.03, 0.06,0.1$
and $M_{h_{t}}=200,400$ GeV, respectively. The numerical results
of the cross sections are shown in Figs.\ref{fig:fig3} and
\ref{fig:fig4}.

The Figs.3 and 4 show the plots of the hadronic cross sections of
the $t\bar{q}$ production as a function of $M_{\Pi_t}$ at the
Tevatron and the LHC, with fixed $\varepsilon=0.06$ and
$M_{h_t}=200, 400$ GeV, respectively. It is shown that the cross
sections of all the processes are at the level of a few pb at the
LHC, which are at least one order of magnitude larger than those
at the Tevatron. On the other hand, we find that there exists the
difference about the cross sections between the LHC and Tevatron.
At the LHC, the cross section of the $t\bar{d}$ production is
larger than that of the $t\bar{u}$ production, but at the
Tevatron, the situation is inverse. The reason is that all the
contributions arise from sea quarks at the LHC but the value
$\bar{u}$-quark gives much more contribution at the Tevatron.

\section{The $tg$ production}

\hspace{0.6cm}With the one-loop FC coupling $tcg$, the $tg$
production can be realized via charm-gluon collision as shown in
Fig.2. The s-channel amplitude is
\begin{eqnarray}
M_{B}&=&\frac{im_{t}^{2}tan^{2}\beta}{16\pi^{2}\upsilon_{w}^{2}}g_{s}^{2}(1-\varepsilon)
\sqrt{2\varepsilon-\varepsilon^{2}}(T_{ij}^{a}T_{jk}^{b})G(p_{3}+p_{4},m_c)\nonumber\\
&&\cdot \{\hspace*{0.3cm}\{-2B_{1}(-p_{3},m_{b},M_{\Pi_{t}})-B_{1}(-p_{3},m_{t},M_{\Pi_{t}})-B_{1}(-p_{3},m_{t},M_{h_{t}})\nonumber\\
&&\hspace*{1.0cm}-B_{0}(-p_{3},m_{t},M_{\Pi_{t}})+B_{0}(-p_{3},m_{t},M_{h_{t}})\nonumber\\
&&\hspace*{1.0cm}+B_{0}(-p_{3}-p_{4},m_{t},M_{\Pi_{t}})-B_{0}(-p_{3}-p_{4},m_{t},M_{h_{t}})\nonumber\\
&&\hspace*{1.0cm}+2m_{b}^{2}C_{0}-4C_{24}-2C'_{24}-2C^*_{24}\nonumber\\
&&\hspace*{1.0cm}-m_{t}^{2}(2C_{22}+C'_{22}-C'_{0}+C^*_{22}-C^*_{0})\hspace*{0.3cm}\}\nonumber\\
&&\hspace*{3.5cm}\overline{u}_{t}(p_{3})\eslash^b(p_{4})(\pslash_{3}+\pslash_{4})\eslash^a(p_{1})Ru_{c}(p_{2})\nonumber\\
&&\hspace*{0.3cm}+2m_{t}[p_{3}\cdot e^b(p_{4})](2C_{22}+C'_{22}+C'_{12}+C^*_{22}-C^*_{12})\nonumber\\
&&\hspace*{3.5cm}\overline{u}_{t}(p_{3})(\pslash_{3}+\pslash_{4})\eslash^a(p_{1})Ru_{c}(p_{2})\nonumber\\
&&\hspace*{0.3cm}+[2(C_{23}+C_{12})+(C'_{22}+C'_{12})+(C^*_{22}+C^*_{12})]\nonumber\\
&&\hspace*{3.5cm}\overline{u}_{t}(p_{3})\pslash_{4}\eslash^b(p_{4})\pslash_{3}(\pslash_{3}+\pslash_{4})
\eslash^a(p_{1})Ru_{c}(p_{2})\nonumber\\
&&\hspace*{0.3cm}+m_{t}(2C_{23}+C'_{23}-C'_{0}+C^*_{23}+C^*_{0})\nonumber\\
&&\hspace*{3.5cm}\overline{u}_{t}(p_{3})\eslash^b(p_{4})\pslash_{4}(\pslash_{3}
+\pslash_{4})\eslash^a(p_{1})Ru_{c}(p_{2})\}
\end{eqnarray}
In the amplitude $M_B$, the three-point standard functions are
defined as
$$C_{ij}=C_{ij}(-p_{4},-p_{3},m_{b},m_{b},M_{\Pi_{t}}),$$
$$C'_{ij}=C_{ij}(-p_{4},-p_{3},m_{t},m_{t},M_{\Pi_{t}}),$$
$$C^*_{ij}=C_{ij}(-p_{4},-p_{3},m_{t},m_{t},M_{h_{t}}).$$

The t-channel amplitude is
\begin{eqnarray}
M_{C}&=&\frac{im_{t}^{2}tan^{2}\beta}{16\pi^{2}\upsilon_{w}^{2}}g_{s}^{2}(1-\varepsilon)
\sqrt{2\varepsilon-\varepsilon^{2}}(T_{ij}^{a}T_{jk}^{b})G(p_{3}-p_{1},m_c)\nonumber\\
&&\{\hspace*{0.3cm}\{-2B_{1}(-p_{3},m_{b},M_{\Pi_{t}})-B_{1}(-p_{3},m_{t},M_{\Pi_{t}})-B_{1}(-p_{3},m_{t},M_{h_{t}})\nonumber\\
&&\hspace*{1.0cm}-B_{0}(-p_{3},m_{t},M_{\Pi_{t}})+B_{0}(-p_{3},m_{t},M_{h_{t}})\nonumber\\
&&\hspace*{1.0cm}+B_{0}(p_{1}-p_{3},m_{t},M_{\Pi_{t}})-B_{0}(p_{1}-p_{3},m_{t},M_{h_{t}})\nonumber\\
&&\hspace*{1.0cm}+2m_{b}^{2}C_{0}-4C_{24}-2C'_{24}-2C^*_{24}\nonumber\\
&&\hspace*{1.0cm}-m_{t}^{2}(2C_{22}+C'_{22}-C'_{0}+C^*_{22}-C^*_{0})\hspace*{0.3cm}\}\nonumber\\
&&\hspace*{3.5cm}\overline{u}_{t}(p_{3})\eslash^a(p_{1})(\pslash_{3}-\pslash_{1})\eslash^b(p_{4})Ru_{c}(p_{2})\nonumber\\
&&\hspace*{0.3cm}+2m_{t}[p_{3}\cdot e^a(p_{1})](2C_{22}+C'_{22}+C'_{12}+C^*_{22}-C^*_{12})\nonumber\\
&&\hspace*{3.5cm}\overline{u}_{t}(p_{3})(\pslash_{3}-\pslash_{1})\eslash^b(p_{4})Ru_{c}(p_{2})\nonumber\\
&&\hspace*{0.3cm}-[2(C_{23}+C_{12})+(C'_{22}+C'_{12})+(C^*_{22}+C^*_{12})]\nonumber\\
&&\hspace*{3.5cm}\overline{u}_{t}(p_{3})\pslash_{1}\eslash^a(p_{1})\pslash_{3}(\pslash_{3}-\pslash_{1})\eslash^b(p_{4})Ru_{c}(p_{2})\nonumber\\
&&\hspace*{0.3cm}-m_{t}(2C_{23}+C'_{23}-C'_{0}+C^*_{23}+C^*_{0})\nonumber\\
&&\hspace*{3.5cm}\overline{u}_{t}(p_{3})\eslash^a(p_{1})\pslash_{1}(\pslash_{3}-\pslash_{1})\eslash^b(p_{4})Ru_{c}(p_{2})
\}
\end{eqnarray}
In the amplitude $M_C$, the three-point standard functions are:
$$C_{ij}=C_{ij}(p_{1},-p_{3},m_{b},m_{b},M_{\Pi_{t}}),$$
$$C'_{ij}=C_{ij}(p_{1},-p_{3},m_{t},m_{t},M_{\Pi_{t}}),$$
$$C^*_{ij}=C_{ij}(p_{1},-p_{3},m_{t},m_{t},M_{h_{t}}).$$

We can use the same way as that for $t\bar{q}$ productions to
obtain the hadronic cross section of $tg$ production.

\newpage
\begin{figure}[thb]
\begin{center}
\includegraphics [scale=0.8]{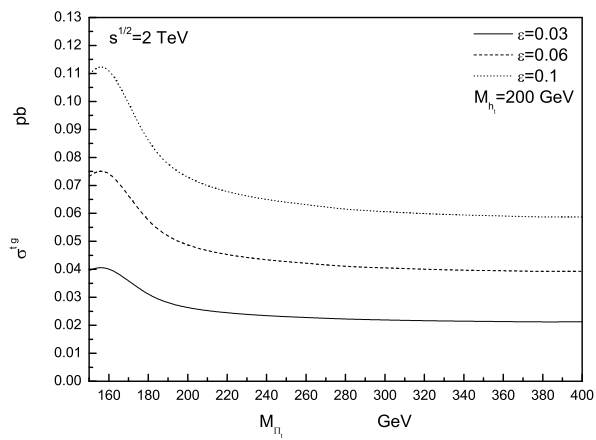}
\includegraphics [scale=0.8]{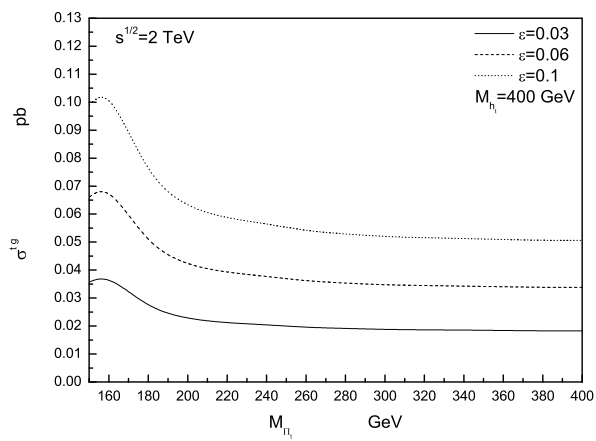}
\caption{The hadronic cross section of $tg$ production as a
function of $M_{\Pi_t}$ at the Tevatron, with $M_{h_t}=200, 400$ GeV, respectively.} \label{fig:fig5}
\end{center}
\end{figure}

\newpage
\begin{figure}[thb]
\begin{center}
\includegraphics [scale=0.8] {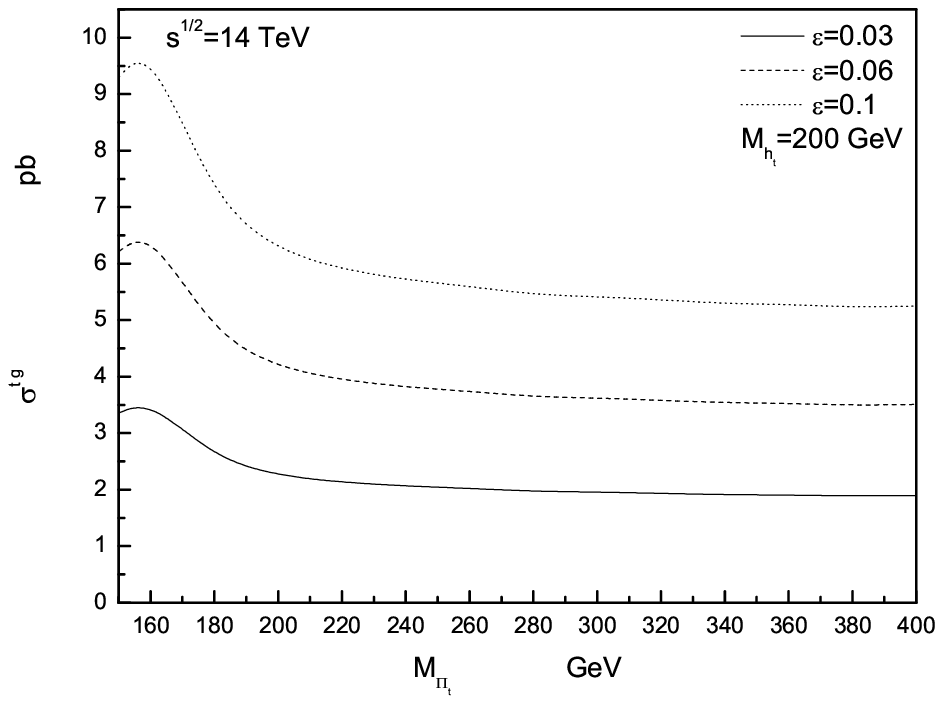}
\includegraphics [scale=0.8] {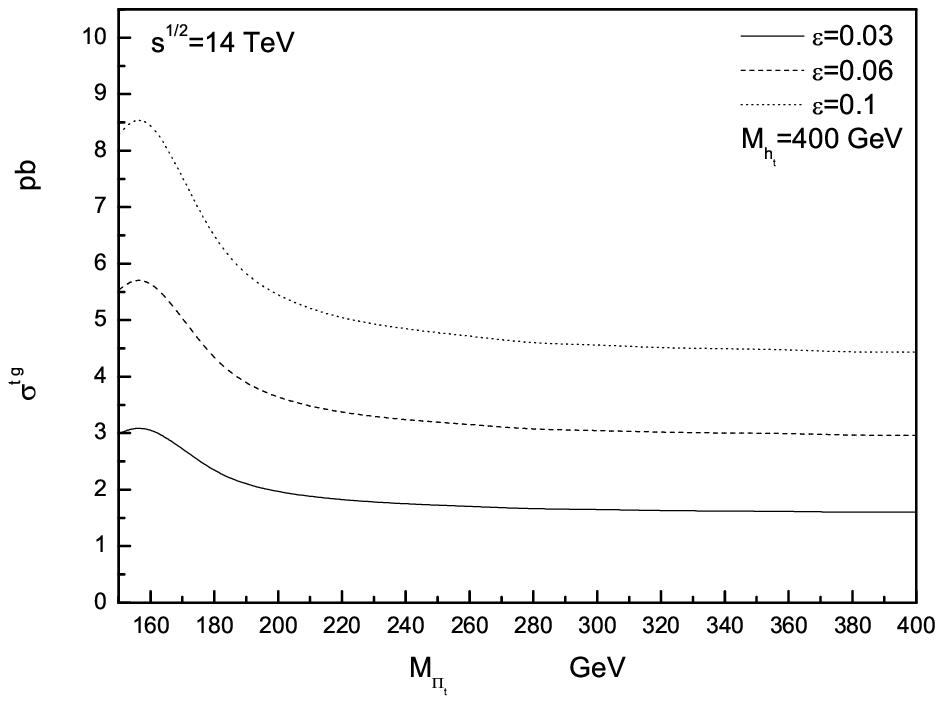}
\caption{The hadronic cross section of $tg$ production as a
function of $M_{\Pi_t}$ at the LHC, with $M_{h_t}=200, 400$ GeV, respectively.} \label {fig:fig6}
\end{center}
\end{figure}

In Figs. 5 and 6, we show the hadronic cross section of $tg$
production varies with $M_{\Pi_{t}}$ at the Tevatron and LHC. The
cross section of $tg$ production is very small at the Tevatron
which is at the order of $10^{-2}$ pb in most parameter spaces.
However, the cross section can reach the level of a few pb at the LHC.

\section{The discussions and conclusions}

\hspace{0.6cm}It is well known that the t-quark decays to $W^{+}b$
with the almost $100\%$ branching ratio, and the t-quark decay
$t\rightarrow W^+b\rightarrow bl\nu_l(l=e, \mu)$ is easier to be
identified than the pure hadronic mode. So, it is suggested that
$bl\nu_l+j$ can be chosen as the search signal. Where $j$ is a
light parton jet. i.e., the search signal should be an energetic
charged lepton, missing $E_T$, a b-quark jet from the t-quark
decay, and a light jet.

For the signal $bl\nu_l+j$, the major source of background is
$W+jj$, where the jets are light quarks or gluons. Since the
signal involves one b-quark in the final state, b-tagging can
effectively reduce the $Wjj$ background and the b-tagging
efficiency is about $60\%$ at the Tevatron Run II and the LHC.
There might be $1\%$ of non-b-quarks misidentified as b-quark,
which gives a suppression of the $W+jj$ background by a factor of
$\approx 50$\cite{IN-TX}. The other significant backgrounds are
the SM single t-quark production where one light quark accompanies
the t-quark in the final state, and the SM $Wb\bar{b}$ production.
Backgrounds with two b-quarks in the final state will mimic our
signal if one of the b-quarks is missed by the b-tagging. On the
other hand, $M_{bW}$, the invariant mass of the W and b-quark,
should be peak at $m_t$ for the signal. Therefore,
 to further eliminate the background, one can impose a constraint
on $M_{bW}$. $M_{{bW}}$ can be experimentally determined by the
reconstruction of t-quark momentum. To find other ways to suppress
the background, one should measure the kinematic distributions and
apply some cuts for the final states. The detailed background
analysis is beyond our work in this paper. Here, we use the
criterion $N_s\geq3\sqrt{N_s+N_B}$£¬ approximately corresponding
to a $95\%$ confidence level, to determine the value of the
background below which the signal should be observable. To
determine the signal $N_s$, one should times the branching ratio
2/9 for the t-quark semileptonically decaying to $e,\mu$. We can
safely assume $N_s\sim 10^5$ with the cross section $1$ $pb$ and
the yearly luminosity $100 fb^{-1}$. With a $95\%$ confidence
level, the order of magnitude of the background should be smaller
than $10^4$ fb. The main background $Wjj$ can be effectively
reduce to the level of $10^3$ fb with b-tagging. So, the LHC has
the considerable capability to observe the signal of the single
t-quark productions via the FC processes. But at the Tevatron, the
number of the signal is too small to be observable.

On the other hand, there is the difference between the production
modes $t\bar{q}(q=u,d,s), tg$ and $t\bar{c}$. $t\bar{c}$ can not
only be produced via the processes involving the loop-level FC
coupling $tcg$, but also be produced via tree-level processes
involving the $\Pi_t^0t\bar{c}$ and $\Pi_t^+c\bar{b}$. The
production modes $t\bar{q}, tg$ only involve the coupling $tcg$
and the changing of their cross sections only depends on the
coupling $tcg$. Therefore, the single t-quark production modes
$t\bar{q}, tg$ have the advantage to study the FC coupling $tcg$
which includes the important information of the TC2 models.

In conclusions, the TC2 model is an interesting dynamical theory
which can explain the EWSB and solve  the heavy t-quark problem.
There exist the tree-level FC couplings in the TC2 model which can
result in the loop-level FC coupling $tcg$. The single t-quark can
be produced via some FC processes involving $tcg$ at the hadron
colliders. In this paper, we study some interesting single t-quark
processes $pp(p\bar{p})\rightarrow t\bar{q},tg$. We find that the
cross sections of these processes can reach the level of a few pb
at the LHC. The LHC running in 2007 will open an ideal window to
study these single t-quark production processes and to obtain some
useful information of the FC couplings resulted by the TC2 model,
furthermore, to probe the EWSB.

\section*{Acknowledgments}

\hspace{0.6cm}This work is supported  by the National Natural
Science Foundation of China under Grant No.10375017, 10575029 and
10575052.

\end{document}